\documentclass[12pt]{iopart}%     Klassen: scrartcl, scrreprt, scrbook iopart
 
\usepackage{graphicx}
\usepackage{picins}
\usepackage{multirow}
\usepackage{multicol} 
\usepackage{tabularx}
\usepackage{amsmath}
\usepackage{amssymb}
\usepackage{lscape}

\usepackage{float} 
\usepackage{exscale}
\usepackage{trfsigns}
\usepackage{txfonts}
\usepackage{units}
\usepackage{listings}  
\usepackage{xspace}
\usepackage{booktabs}

\usepackage[
	table % Load for using rowcolors command in tables
]{xcolor}

\newcommand{\SEpunkt}{\dot{S}_E}
\newcommand{\kb}{k_B}
\newcommand{\intnull}{\int\limits_0^\infty}
\newcommand{\intdelta}{\int\limits_{\Delta E}^\infty}
\newcommand{\Ppunkt}{\dot{P}(E,t)}
\newcommand{\tp}{\tilde P}
\newcommand{\peins}{P^{(1)}}

\usepackage[ngerman, english]{babel} 
\usepackage[T1]{fontenc}
\usepackage[latin1]{inputenc}

\usepackage[automark, nouppercase]{scrpage2}

\begin{document}
 \title{Inelastic collisions as a source of entropy?}
 \author{J\"{u}rgen Schlitter}
\address{ Lehrstuhl f\"{u}r Biophysik\\
           Ruhr-Universit\"{a}t Bochum ND 04, \\
          44780 Bochum, Germany}
    \ead {\href{mailto:juergen.schlitter@rub.de}{juergen.schlitter@rub.de}}
        
 \date{2011-04-07}

\setlength{\parindent}{5mm}
\setlength{\parskip}{1.5mm }
\begin{abstract}
Activation and deactivation by inelastic collisions have been extensively studied at
unimolecular reactions in gas phase where they are crucial for equilibration. As equilibration
means an increase of entropy, the mechanism can also be considered responsible for entropy
production. The numerous results of theoretical treatments show a remarkable agreement
with experimental data. Analysis of the statistical physicochemical argumentation reveals the
stochasticity of quantum transitions as the essential assumption. Under this premise, master
equations have been used that are known to deliver equilibria and entropy production.\\
\indent Here we examine the hypothesis that the ubiquitous inelastic interactions in gas and liquid
phase may represent a source of entropy beyond chemical reactions, rotational activation/deactivation
being the prevailing mechanism in gas dynamics at room temperature. For a quantum mechanical
two-state model the master equations are formulated which yield entropy production and equilibration
in translational degrees of freedom until, at conserved energy, a stationary Maxwell-Boltzmann
distribution is reached.\\
\indent The relaxation rates show features that can be checked by monitoring thermal relaxation in gas phase. Depending on the composition, first
or second order processes are predicted. The temperature dependence of relaxation rates is
determined by the activation energy of the dominating quantum transition. Thus, experimental verification
will allow to decide to which extent this hypothesis describes thermal relaxation. It would support a connection between the macroscopic second law of thermodynamics and the microscopic stochastic collapse in quantum mechanics. They are both experimentally secured facts which in theory emerge as special elements
beyond Hamiltonian dynamics.\\
\indent It is also shown that inelastic collisions are connected with velocity-dependent forces and result
in a new analog of the Fokker-Planck equation where they replace Langevin dynamics as
a non-Hamiltonian dissipative mechanism.\\

%\nocite{*}

\noindent{\it Keywords\/}: Entropy production, dissipation, Fokker-Planck equation, master equation, inelastic collisions, gas dynamics \\

\noindent{\it PACS\/}: 05.20.-y Classical statistical mechanics\\ 
\hspace*{0.9cm} 02.50.-r Probability theory, stochastic processes, and statistics\\

\end{abstract}
\pagebreak
\section{Introduction}

Recently, several new attempts were made to prove second-law conform behavior of classical systems when work is applied \cite{jarzinski,kawai,parrondo} or at sudden release of a constraint \cite{schlitter}. It is in fact possible to derive an increase of entropy in such situations using classical mechanical statistics. On the other hand, it is known that Gibbs entropy - which is always referred to here - is time-invariant under classical Hamiltonian dynamics and cannot change in the situations mentioned. The seeming contradiction is resolved by the observation that the above theories are all based on the assumption of equilibria, which are known to be states of maximum entropy \cite{jaynes}. For instance, in our own approach it is proven that immediately after release of a constraint, any system is found in a state of higher entropy - provided it is in equilibrium at the start and the new position \cite{schlitter}. However, when it is assumed that a process ends with equilibrium, the treatment tacitly departs from the field of Hamiltonian dynamics without declaring the new theoretical basis. One has to conclude that the cited approaches are formally correct, but the physical mechanism of entropy production during equilibration is still the ultimate unsolved problem. Moreover, it has to be described using an adequate non-Hamiltonian ansatz. An example of this kind is the derivation of Jarzynski's work theorem from Langevin dynamics, i.e. a non-Hamiltonian phenomenological ansatz \cite{seifert}. \\ 
\indent Because of the maximum entropy property of equilibria, relaxation mechanisms are candidates for the explanation of entropy production. Equilibration by activating and  deactivating inelastic collisions were extensively studied at unimolecular reactions in gas phase \cite{atkins,forst,pritchard}. These are reactions of the type
\begin{eqnarray}
\label{reaction}
M + A\rightleftharpoons  M + A^* \rightarrow\ M + P
\end{eqnarray}
where molecules \textit{A} are activated/deactivated by collisions with molecules \textit{M}, and can decay from the activated state \textit{A*} to product(s) \textit{P}. \textit{A} and \textit{M} may be the same molecular species. The last step is responsible for the denomination as a unimolecular reaction. It requires 'internal energy randomization'\cite{pritchie} for transferring energy to the crucial reaction coordinate. The results of numerous theoretical treatments show a remarkable agreement with experimental data. Analysis of the statistical physicochemical argumentation reveals the stochasticity of quantum transitions as the essential assumption. Under this premise, master equations have been used that are known to deliver equilibria and entropy production. Rotational and vibrational activation were considered for the step $A \rightleftharpoons A^*$ with experimentally determined inelastic cross sections \cite{pritchard}. The final decay also needs deactivating collisions as indicated by pressure dependence of the transmission coefficient of the decay which can increase with pressure due to increasing frequency of collisions \cite{laidler}. \\
\begin{figure}[ht]
\begin{center}
\fbox{
\begin{minipage}{0.65\textwidth}
      	\includegraphics[width=\textwidth]{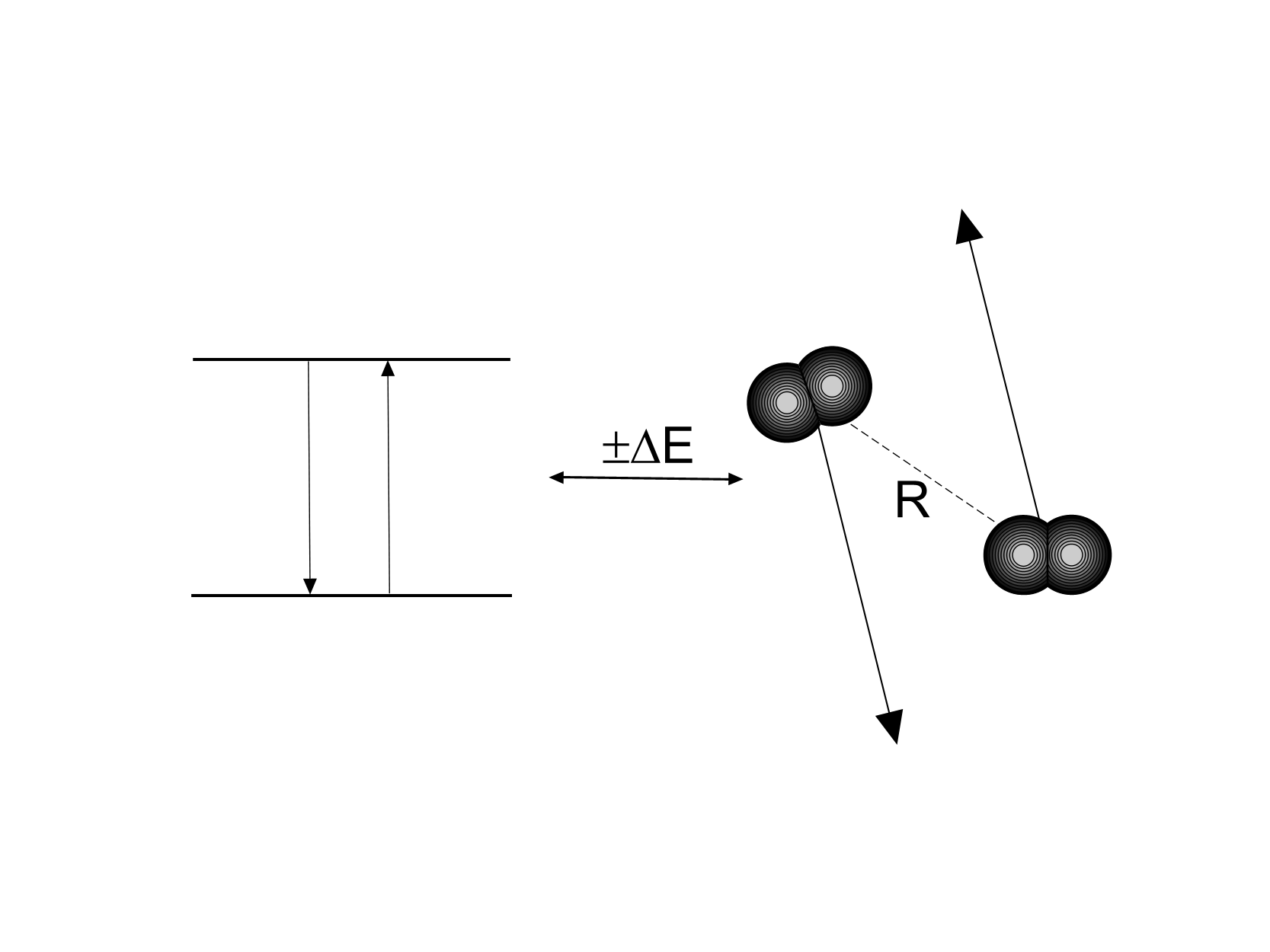}
 \end{minipage}}
	\caption[]{\label{fig1}The principle of inelastic QM/MM interaction. At collision of molecules, energy transfer between translational and internal degrees of freedom by inelastic processes can occur, which are the focus of this investigation.} 
\end{center}
\end{figure}
\indent The successful description of inelastic collisions as stochastic events by master equations can be regarded as an approximation that requires justification by invoking decoherence phenomena \cite{zurek1}\cite{zurek2}. Here we take stochasticity for granted. We reconsider the old experiments and theories as they suggest a very general role of such collisions. The proper hypothesis of this work claims that the ubiquitous inelastic collisions in gases and liquids provide an entropy-producing mechanism already under the threshold of chemical reactions. As will be shown, there is a realistic way for experimental verification.\\
\indent The hypothesis of entropy-producing inelastic collisions is substantiated by the existence of QM energy levels that can be excited even at low and medium temperatures. The proper verification/falsification of the hypothesis requires prediction of measurable features. This is done by theoretical treatment of an internal two-state model which is shown schematically in Fig. \ref{fig1}. It possesses the essential features of  relaxation mechanisms employed elsewhere \cite{forst,pritchard} and allows to derive relaxation rates.\\
\indent Theories of unimolecular reactions focused on the equilibration among molecular species, i.e. chemical relaxation. Our hypothesis concerns thermal relaxation of both internal and translational degrees of freedom of the colliding molecules, which were not considered at unimolecular reactions. Translations are denoted and treated as classical (or molecular mechanical 'MM') while the QM degree of freedom are represented by two distinct energy levels. We investigate the impact of the stochastic QM transitions on the MM system of translations. There are also theories which take into account the influence of a stochastic bath on a QM system, e.g. by applying a stochastic Liouville equation \cite{kubo,freed} to spins or considering quantum information \cite{schulz}. In this study, however, the QM system is the 'stochastic bath' for classically moving particles.\\
\indent The article is organized as follows: In section 2, we examine available inelastic processes which are neglected or inadequately treated in usual classical dynamics. The master equations of the two-state model are developed and transformed to equations for various entropies in section 3 where also rate constants are derived. In the following section 4 we propose incorporating the inelastic process in the Fokker-Planck equations as a dissipative mechanism which can replace the phenomenological Langevin dynamics. In the final discussion, the focus is placed on the justification of the underlying assumptions and formalism, and the relation to Shannon information.\\

\section{Inelastic processes}

\indent When two neutral molecules collide without chemically reacting as shown in Fig.\ref{fig1}, part of the kinetic energy $E_1 + E_2$ of the centers of mass is transiently transformed into internal energy $E_{int}$, for instance, by induced polarization. Here, internal energy denotes all kinds but translational energy. The resulting van der Waals interaction is the only effect at slow collisions below the threshold of excitations. Persistent exchange of energy with rotational or vibrational levels takes place when molecules are in excited states or kinetic energy is large enough to allow excitation. A quantum mechanical calculation proves that internal energy remains unchanged only when the collision is very fast, the so-called rapid-passage case \cite{Messiah}.\\
\indent Here, internal energy comprises all quantized degrees of freedom with discrete spectra including rotations. At room temperature 300 K corresponding to a wave number \mbox{$\upsilon (k_BT)\approx 200cm^{-1}$,} rotations ($\upsilon_{R}\leq 20cm^{-1}$ ) play a major role in gases and liquids. In liquids, also librations \mbox{($\upsilon_{L}\approx 300-900cm^{-1}$ in $H_2O$ at $0  C^\circ$)} and hindered translations \mbox{($\upsilon_{L}\approx 200cm^{-1}$ in $H_2O$)} contribute to energy exchange at collision. Monatomic and most homonuclear diatomic molecules are rotationally inactive and provide no suitable internal degrees of freedom at 300 K. Here, only impurities and collisions with walls confining the gas present mechanisms for inelastic energy exchange. Moreover, vibrations are available with low frequencies at larger molecules, and at higher temperature electronic transitions are induced everywhere by collisions. The variety of low energy transitions available in gases and liquids that can be excited at room temperature demonstrates the importance of inelastic interaction in the classical regime. As a consequence, a finite fraction of interacting particles will, after a while, be found in some excited state, and a permanent uptake and release of energy by inelastic processes is taking place. This side-effect of intermolecular interaction and its implications are the subject of this study.\\
\indent It is worth considering the reason why entropy is invariant when the dynamics is governed by a Hamiltonian $H(q,p)=K(q,p)+V(q)$, (q,p) being positions and momenta. For the classical Gibbs entropy
\begin{eqnarray}
\label{gibbsentropy}
S=-k_B \int\limits_{-\infty}^\infty P(q,p,t)\; ln P(q,p,t)\; dq\; dp
\end{eqnarray}

the time derivative can be written
\begin{eqnarray}
\label{from}
\dot{S}=-k_B \int\limits_{-\infty}^\infty {\dot{P}(ln P +1)\;dq\; dp}
\end{eqnarray}
\begin{eqnarray}
\label{secondform}
=k_B \int\limits_{-\infty}^\infty P \; \frac{\partial F}{\partial p}\; dq\; dp
\end{eqnarray}
$k_B$ ist the Boltzmann constant and \textit{P} the probability density. The second form (\ref{secondform}) obtained by partial integration proves that only a momentum-dependent force $\dot{p}=F(q,p)$ could influence entropy, but not a force derived from a potential function $V(q)$ as assumed in Hamiltonian dynamics. There also the energy $H(q,p)$ is invariant. On the other hand, inelastic interactions stand for uptake or release of energy portions $\Delta E$ in short time intervals $\Delta t$. Because kinetic energy changes according to $dE = vdp = vFdt$, one finds that during the transition a force $F$ 
\begin{eqnarray}
F = \frac{1}{v} \frac{\Delta E}{\Delta t}
\end{eqnarray}
is acting. The significance of this relation lies in the fact that here the force indeed depends on velocity $v$ or momentum $p$. This is a strong indication for entropy production by inelastic processes that can be extended to a mathematical argument. Here, however, the final proof will be given by a derivation that starts from (\ref{from}). Velocity-dependent forces are characteristic of Langevin dynamics. In simulations, they occur in thermostat algorithms where they mimic heat exchange with an environment.\\

\section{Master Equations of a two-state model}

\begin{figure}[ht]
  \begin{center}
\fbox{
\begin{minipage}{0.65\textwidth}
      	\includegraphics[width=\textwidth]{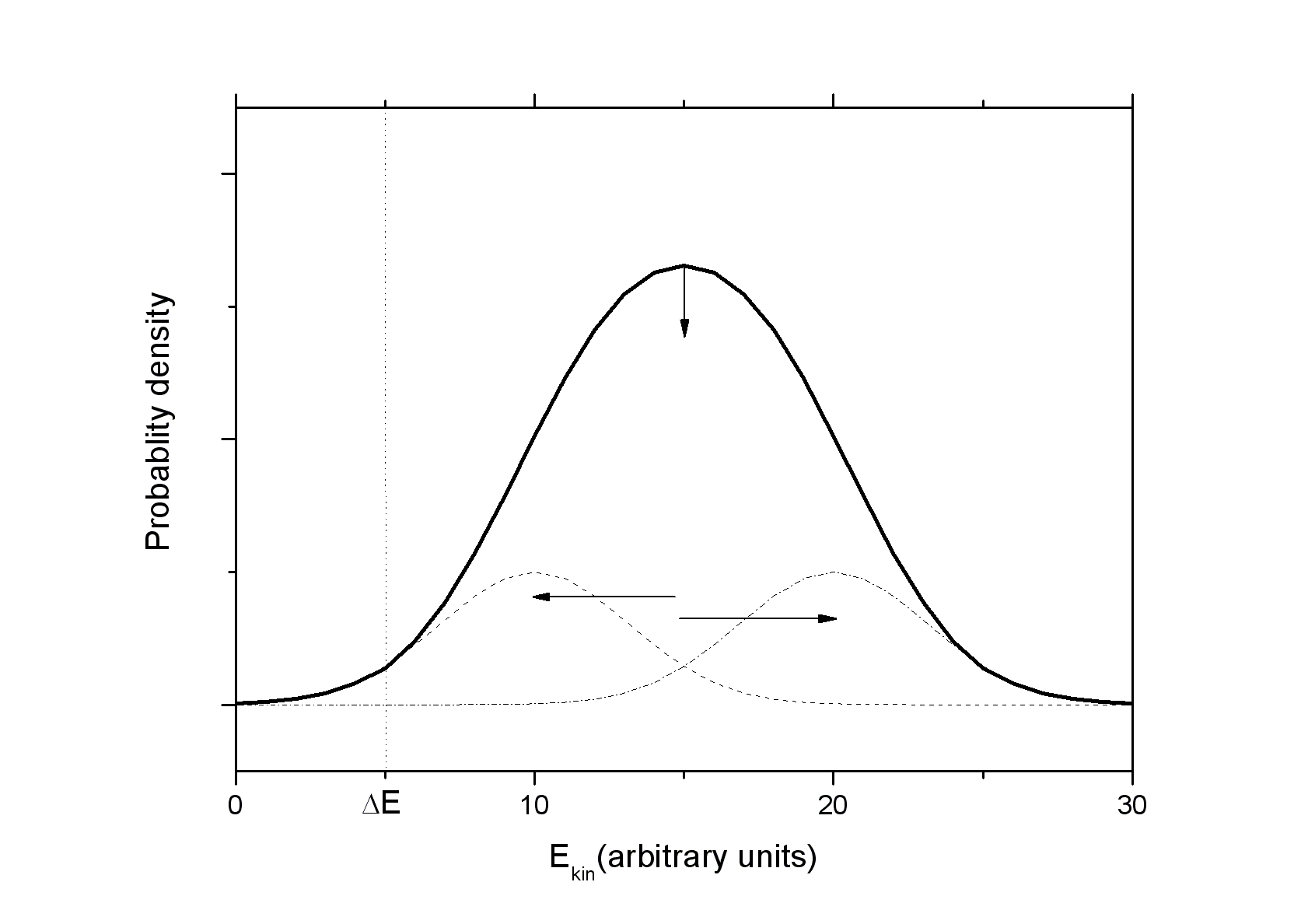}
 \end{minipage}}
   		\caption[]{\label{seealso}Broadening of probability density by release (left) and uptake (right) of energy $\Delta E$. Fractions of the PDF are shifted to the flanks whereas the original PDF (bold) is flattening. Note that systems with $E< \Delta E$ cannot give away their energy. They are not damped, but can be speeded up.}   	
   \end{center}
\end{figure}

We now consider a two-state model according to Fig.\ref{fig1} The following assumptions are made: (i) Atoms and molecules are point particles with classical probability density functions (PDF) $P^{(1)}(\textbf{x},\textbf{p},t)$ for a single particle and $P^{(N)}(\textbf{x},\textbf{p},t)$  for an N-particle system together with the corresponding Gibbs entropy (\ref{gibbsentropy}). $(\textbf{x},\textbf{p})$ are the respective Cartesian coordinates and momenta of the centers of mass. A classical treatment is justified by the uncertainty principle which predicts classical behavior for a box length $\Delta x /nm >> (M/ (g/mol))^{-1}$, i.e. already in the higher nanometer range at any molecular mass $M$. (ii) In the classical world QM transitions can be taken into account as forces acting for a period of time or by discrete energy portions when they couple to translation. As a matter of fact, molecules and single atoms in ordinary experiments at room temperature are at the interface between the classical MM and QM regime: their rotational or electronic degrees of freedom require a QM treatment. (iii) The transitions are random, stochastic processes, which justifies a master-equation approach, and (iv) energy exchange is assumed to occur strictly without net change in energy on the average and therefore without transfer of entropy.\\
We neglect the simultaneous energy transfer on both colliding particles. Instead, we consider the action of a generic, non-specified QM system on single particles that feeds in energy packets (power strokes) $+\Delta E >0$ or removes energy in portions $-\Delta E$. The symbol $\Delta E\equiv |\Delta E|$  stands for the absolute value, $E=E_{kin}=p^2/2m$ is the kinetic energy of translation. As the events of energy exchange are random, stochastic QM transitions we employ a master equation for the classical probability density function to account for their action on translation. Elastic interactions are not included from the beginning, but added later. We start with the development of the energy density $P(E,t)$, $0\leq E < \infty$, which facilitates the connection with the QM system.\\ 
\indent Let us assume that the uptake of $\Delta E$ occurs with rate $w^+$ and the release of the same amount of energy with rate $w^-$. Uptake of energy simply means that in a time interval $dt$, part of the density is lost everywhere and shifted to higher energy, where it appears at  some $E > \Delta E$,
\begin{align}
\label{upt}
\nonumber
&dP_{upt}(E,t)=(-w^+P(E,t)+w^+P(E-\Delta E,t))\;dt  &\text{if $E> \Delta E$}\\ 
&dP_{upt}(E,t)= -w^+P(E,t)\;dt    &\text{if $E < \Delta E$}
\end{align}
A release of $-\Delta E$ takes place only if the energy is available, i.e. for $E> \Delta E$. Therefore, one has two cases
\begin{align}
\label{rel}
\nonumber
dP_{rel}(E,t)=(-w^-P(E,t)&+w^-P(E+\Delta E,t))\;dt  &\text{if $E> \Delta E$}\\ 
dP_{rel}(E,t)= \hspace{2.2cm} &+w^-P(E+ \Delta E,t)\;dt    &\text{if $E < \Delta E$}
\end{align}
In any case, part of the density is shifted to lower energy during release, see also Fig. \ref{seealso}.\\
\indent By integration over energy it is easily seen that the normalization of probability density is always maintained. Taking both processes together and making use of the Heaviside step function $\Theta = \Theta (E-\Delta E)$, one obtains
\begin{equation}
\label{oneobtains}
\begin{split}
dP(E,t)&=\lbrack-(w^{+}+\Theta w^{-})P\,(E,t)+\Theta w^{+}P\,(E-\Delta E,t)\\ \quad &+w^{-}P\,(E+\Delta E,t)\rbrack\;dt
\end{split}
\end{equation}
Because of the different behavior at low energy in (\ref{upt}-\ref{rel}), the rates $w^+$ and $w^-$ cannot be equal, but differ by a factor $\varphi = w^-/w^+$. It is assumed that there is no net change of energy, i.e.
\begin{eqnarray}
\label{3.4}
w^- \int\limits_{\Delta E}^\infty P(E,t)\,dE=w^+ \int\limits_0^\infty P(E,t)\,dE
\end{eqnarray}
which implies $w^+\leq w^-$ and $\varphi= \int\limits_{\Delta E}^\infty P(E,t)\,dE$. This condition will be employed in the appendix for calculating the time dependency of entropy at constant energy.\\
\indent It is obvious that this dynamics is suited to broaden an initially narrow distribution or constrict a broad distribution, and possibly produce entropy. This is rigorously proven in the appendix for the time derivative of the entropy $S_E$ belonging to $P(E,t)$, 
\begin{eqnarray}
S_E=-k_B \int\limits_0^\infty P(E,t)\ln P(E,t)\,dE
\end{eqnarray}
and, more importantly, for one-particle phase-space Gibbs entropy $S^{(1)}$ defined by
\begin{eqnarray}
S^{(1)}=-k_B \int\limits_0^\infty P^{(1)}(\textbf{p},t)\ln P^{(1)}(\textbf{p},t)\,d^3p
\end{eqnarray}
of the one-particle phase-space PDF $P^{(1)}(\textbf{p},t)$  on the assumption of energy conservation. The result is $dS^{(1)}/dt\geq 0$, i.e. entropy never decreases. More precisely, entropy turns out to increase as long as the stationary state is not reached where entropy remains constant. In other words, coupling to and energy exchange with a quantum system is causing dissipation and equilibration in a classical system.\\
\indent The stationary solution is obtained in the same way (see appendix) or by solving eq. \mbox{(\ref{upt}-\ref{rel})} with the condition $dP\equiv dP_{upt}+dP_{rel}=0$. The unique integrable, real solution of either approach is a decreasing exponential which is actually independent of $\Delta E$ although, according to (\ref{upt}-\ref{rel}), the two intervals have to be treated separately.  It is also independent of the transition rates $w^{+/-}$, which confirms its equilibrium character. A posteriori, one can write it as a Maxwell-Boltzmann distribution
\begin{eqnarray}
\label{togetherwith}
P_s^{(1)}(E)=\beta \exp{(-\beta E)}
\end{eqnarray}
with $\beta= 1/k_B T$ which is characterized by mean energy or temperature $T$. This finally proves that stochastic interaction with a QM system is an entropy-producing physical mechanism that generates the well-known equilibrium distribution $P_S(E)$ of maximum entropy. This also holds for the one-particle PDF $P_s^{(1)}(p)$, see appendix. The mechanism proposed here is thus equivalent to the phenomenological Langevin mechanism that is usually assumed to be present and actually causes convergence towards equilibrium.\\
\indent According to eq. \ref{A6}, entropy production and equilibration occur with a rate $w^+=\varphi w^-$, $w^-$ being the rate of energy transfer into the quantum system. Energy conservation (\ref{3.4}) together with (\ref{togetherwith}) reveals  $\varphi$ to be a Boltzmann factor $\varphi=\exp{(-\beta \Delta E)}$, which indicates also equilibrium in the QM system. This finally allows application to inelastic collisions $A + A\rightleftharpoons  A + A^*$. Collision theory predicts a pressure dependence of second order and hence a relaxation rate 
\begin{eqnarray}
\label{3.8}
w^+\propto \sigma_{in} \exp{(-\beta \Delta E)}\,p_A^2
\end{eqnarray}
for a homogeneous gas with pressure $p_A$ if the molecules can be excited/deexcited. If there is a major rotationally inactive component $I$ with pressure $p_I$ that offers no quantum transitions at a given temperature, one expects
\begin{eqnarray}
\label{3.9}
w^+\propto \sigma_{in} \exp{(-\beta \Delta E)}\,p_Ap_I
\end{eqnarray}
where $p_A$ is the pressure of an active impurity and $\sigma_{in}$ the respective inelastic cross section. These relaxation rates possess characteristic properties that can be checked experimentally, namely a typical temperature dependence given by the threshold energy $\Delta E$ of the (lowest) quantum transition and a pressure dependence due to the composition of the gas.  Relaxation is expected to be strongly  hampered at rotationally inactive gases. The two-state model will be valid at sufficiently low temperature where higher states are not excited; it will be approximately valid where such processes take place and a larger relaxation matrix is required. Resonance effects will not alter the essential features of the relaxation rate. It was tacitly assumed that the above inelastic cross section is an average, and likewise that a temperature can be assigned to the system during relaxation.\\

\section{Analog of the Fokker-Planck equation}

The full time-dependence of the PDF $P^{(1)}(\textbf{x},\textbf{p},t)$ is usually described by a Fokker-Planck equation (FPE), assuming that dissipation is given by the Langevin equation of motion, \mbox{$\ddot{\textbf{x}}=-\varsigma \dot{\textbf{x}} + \textbf{A}(t)$,} with friction constant $\varsigma$ and a stochastic force $\textbf{A}(t)$ with vanishing mean. The generalized FPE or Chandrasekhar equation \cite{McQuarrie} is a differential equation of the form 
\begin{eqnarray}
\label{oneparticle}
\dot{P}^{(1)}=\textbf{R}\,P^{(1)}+\textbf{D}\,P^{(1)}
\end{eqnarray}
containing two differential operators, $\textbf{R}=\textbf{v}\cdot \nabla_x +(\textbf{F}(\textbf{x})/m)\cdot \nabla _v $ and $\textbf{D}= \varsigma \nabla _v (\textbf{v}+(k_B T/m)\nabla _v$ with $\textbf{v}=\dot{\textbf{x}}$. $\textbf{R}\,P^{(1)}$ is the regular part following from Hamiltonian dynamics with force $\textbf{F(x)}=-\nabla _x V(\textbf{x})$ derived from an external potential. The term $\textbf{D}P$ being proportional to the friction constant $\varsigma$, is non-Hamiltonian and accounts for dissipation according to the Langevin equation.\\ 
\indent In contrast, we here investigate dissipation basing on inelastic interaction with a QM system.  Equation (\ref{oneobtains}) can be given the shape of an integral operation writing $\dot{P}=\textbf{I}_E\,P$. Then the respective change of $P^{(1)}$ can be written as $\dot{P}_{QM/MM}^{(1)}=\textbf{I}\,P^{(1)}$, where $\textbf{I}$ is the integral operator 
\begin{equation}
\begin{split}
\textbf{I}P^{(1)}(\textbf{x},\textbf{p},t)&=\frac{2}{p}\int\limits_0^\infty \lbrack -(w^+ +\Theta w^-)\delta (E'-E)+ \Theta w^+ \delta (E'-E+\Delta E)\\ \quad &+ w^- \delta (E'-E-\Delta E)\rbrack E'P^{(1)}(\textbf{x},\textbf{p'},t)\,dp'
\end{split}
\end{equation}
Here $E=p^2/2m$ and $E'=(p')^2/2m$. The direction of momentum is not affected. The transition from $P$ to $P^{(1)}$ is derived in the appendix. The operator $\textbf{I}_E$, and hence also $\textbf{I}$, do produce entropy and have a stationary solution which coincides with the Maxwell-Boltzmann distribution. It depends on the rates $w^{+/-}$ and threshold energy $\Delta E$ and can replace the Langevin operator $\textbf{D}$ to account for dissipation. Therefore, we propose a new equation that is formally analog to the generalized FPE $\dot{P}^{(1)}=\textbf{R}\,P^{(1)}+\textbf{D}\,P^{(1)}$ and describes the dissipative part of particle interaction, namely
\begin{eqnarray}
\label{notethat}
\dot{P}^{(1)}=\textbf{R}\,P^{(1)}+\textbf{I}\,P^{(1)}
\end{eqnarray}
It has the same regular part $\textbf{R}\,P^{(1)}$, but a different dissipative part $\textbf{I}\,P^{(1)}$ that is typical of inelastic QM/MM interaction. Note that (\ref{notethat}) does not fully replace the FPE equation because elastic interactions are excluded for the time being. The stationary solution of the new equation is
\begin{eqnarray}
P_S^{(1)}(\textbf{x},\textbf{p})=N_S^{-1}\exp{[-\beta\,(\textbf{p$^2$}/2m +V(\textbf{x}))]}
\end{eqnarray} 
like for the FPE. $N_S$ is a normalization constant. The stationarity with $\textbf{I}$ was proven in the previous section, and is easily seen for $\textbf{R}$ by using the above explicit form of the operator.\\
\indent The new integro-differential equation (\ref{notethat}) for a single particle is straightforwardly generalized to N-particle systems with $\textbf{x}=(\textbf{x}_1 ...\textbf{x}_N), \textbf{p}=(\textbf{p}_1 ...\textbf{p}_N)$ and a density $P^{(N)}(\textbf{x},\textbf{p})$. Defining the operators $\textbf{R}_i$ and $\textbf{I}_i$ as one-particle operators like above which act each only on the phase-space coordinates $\textbf{x}_i$ and $\textbf{p}_i$, one constructs the N-particle operator as a sum $\sum_i (\textbf{R}_i+\textbf{I}_i)$ that governs the motion of the N-particle density as
\begin{eqnarray}
\label{nparticle}
\dot{P}^{(N)}=\sum_i (\textbf{R}_i+\textbf{I}_i)P^{(N)}
\end{eqnarray}
The increase of entropy due to the integral terms $\textbf{I}_i$ follows from the behavior proven for the one-particle density , while the regular parts $\textbf{R}_i$ do not contribute. The stationary solution is the Maxwell-Boltzmann distribution 
\begin{eqnarray}
\nonumber
P_S^{(N)}(\textbf{x},\textbf{p})&=P_S (\textbf{x})\prod_{i=1}^{3N}P_{s,i}^{(1)}(\textbf{p}_i)\\
\nonumber
P_{s,i}^{(1)}(\textbf{p}_i)&= N_i^{-1} \exp{(-\beta \textbf{p}_i^2 /2m_i)}\\
P_S^{(N)}(\textbf{x})&=N_0^{-1}\exp{(-\beta V(\textbf{x}))}
\end{eqnarray}
which is known to be the maximum entropy PDF. The $N_i$ are normalization factors. Here, $V(\textbf{x})$ also includes all elastic interactions between particles and external potentials. They were not yet included in the one-particle equation (\ref{notethat}), but occur only in the N-particle system (\ref{nparticle}).\\ 
\pagebreak
\section{Summary and discussion}

We have formulated the hypothesis that the ubiquitous inelastic collisions in gas and liquid phase represent a source of entropy. Stochasticity of the underlying quantum transitions is the essential assumption to be made. It is supported by the numerous applications to unimolecular reactions where the experiments are well described. Rotational levels are available for excitation at ordinary temperature, followed by vibrations. We have proposed to check the hypothesis experimentally by measuring (or evaluating existing data about) thermal relaxation in gas dynamics for which the variation with temperature, pressure and composition (\ref{3.8},\ref{3.9}) was derived from a model. Note that the dependence on pressure and composition, for instance the presence of inert gases, was already analyzed at unimolecular reactions \cite{pritchard}. Apparently, it is possible to  answer the important question of entropy production experimentally. The optimum temperature range should be low enough  (a) to avoid chemical reactions and (b) to facilitate evaluation by restricting excitation to essentially one single energy level.

\indent It is often believed that elastic collisions of molecules in a gas provide a mechanism for thermal relaxation, i.e. emergence of a Maxwell-Boltzmann distribution independent of the starting conditions. The Boltzmann equation is occasionally invoked as an approximation for dilute gases, and generalized for dense systems \cite{sherman}. In principle, dephasing cannot be excluded to yield an entropy-producing master equation as part of the equation of motion of the density matrix as shown for model systems \cite{zurek2}. On the other hand, quantum mechanical potential scattering is more closely related to its classical counterpart than inelastic scattering, and classical potential scattering is entropy-conserving in arbitrary systems. In the end, the role of elastic scattering is also a question that can be answered by the proposed experiments.		

\indent If there is a stochastic process which interferes with Hamiltonian dynamics and can be described using master equations, then it is known to imply entropy production for mathematical reasons. This can be understood by invoking the effective identity of entropy and (missing or needed) Shannon information. Stochastic events hamper the prediction of the future of a system from its present starting conditions. Each time, extra information has to be added for a full description of the system developing in time. In a discrete representation, the string that fully encodes the system in phase space thus becomes longer and longer. This holds similarly for the PDF with the reservation that information has an upper limit dependent on energy. The second law of thermodynamics clearly states that actually this happens permanently in reality.

\indent A common way of introducing dissipation is the use of phenomenological Langevin dynamics which adds stochastic and velocity-dependent forces to Hamiltonian dynamics. We have shown in passing that in fact stochastic energy transfer by inelastic collisions results in velocity-dependent forces. The corresponding terms can fully replace the Langevin terms in the Fokker-Planck equations to induce dissipation and relaxation to Maxwell-Boltzmann distributions. This has far-reaching consequences for the explanation of numerous relaxation phenomena.

\indent Up to this point, is has become clear that - according to our hypothesis - inelastic collisions are candidates for causing entropy production. Moreover, this suggests that quite generally energy exchange of any QM system with a MM system via discrete transitions as illustrated by Fig.\ref{fig1} will produce entropy in both systems like in the model. What remains open is the question when and why the QM transitions can be treated this way, i.e. as stochastic power strokes for the classical system. This touches the explanation of the transition from QM to classical dynamics, which is out of the scope of this work. It is still under debate and subject to experiments \cite{zurek1}\cite{zurek2}.

\indent The available experiments on unimolecular reactions already suggest two interesting conclusions that could be corroborated by the proposed new experiments. Firstly, they indicate that colliding molecules in large systems - where the reactions are satisfactorily described -  behave like under measurement conditions. Transitions between energy levels involve the collapse of the wave function occurring with a probability that is well described by QM, but the unpredictability of the time point of the collapse makes them a perfect random generator. Secondly, there seems to exist a direct connection between macroscopic entropy production as stated by the second law of thermodynamics and the microscopic stochastic collapse in quantum mechanics. Apparently, stochasticity can explain the increase of  Shannon information or physical entropy. This would connect two experimentally secured facts, quantum transitions and the second law,  which in theory emerge as special elements beyond Hamiltonian dynamics.

\section{Acknowledgment}
\label{sec:ack}

The author thanks Katrin Augustinowski and Martina Bamberg for their help at the preparation of the manuscript.

\appendix

\section{ Time derivative of entropy }

We first derive the time derivative of the entropy $S_E$,
\begin{eqnarray}
\SEpunkt / \kb=- \intnull \Ppunkt\, (\ln P (E,t)+1)\,dE
\end{eqnarray}
using
\begin{equation}
\begin{split}
\Ppunkt &= (-w^+ - \Theta(E-\Delta E)w^-)\underbrace{P(E,t)}_{P}+ w^+ \Theta (E-\Delta E)\underbrace{P(E-\Delta E,t)}_{P^-} \\ & \quad+w^- \underbrace{P(E+\Delta E,t)}_{P^+}
\end{split}
\end{equation}
from (\ref{oneobtains}). The one-particle entropy $S^{(1)}$  will be treated in a second step. As the integral over $P$ was already shown to be conserved, it suffices to consider
\begin{equation}
\begin{split}
\SEpunkt / (\kb w^-)&= \frac{-1}{w^-}\intnull \dot{P}\ln P\, dE\\
& \quad = \varphi \underbrace{\intnull P \ln P\, dE}_A + \underbrace{\intdelta P \ln P \,dE}_B -\varphi \underbrace{\intdelta P^- \ln P\, dE}_C - \underbrace{\intnull P^+ \ln P\, dP}_D
\end{split}
\end{equation}
This is where energy conservation (\ref{3.4}) comes into play. In contrast to $P$ and $P^-$, $P^+$ is not normalized in $(0,\infty)$. The corresponding normalized PDF is $\tilde{P}=P^+ / \varphi$ because energy conservation demands $\intnull P^+ dE=\intdelta P dE= \varphi$. Then one finds
\begin{equation}
\begin{split}
\label{A4}
C&=\intdelta P^- \ln P\, dE= \intnull P \ln P^+ \,dE =\intnull P \ln (\tp \varphi)\,dE\\
&=\intnull P \ln \frac{\tp}{P}\,dE+ \intnull P \ln P \,dE + \ln \varphi\\
&=\intnull P \ln \frac{\tp}{P}\,dE +A + \ln \varphi
\end{split}
\end{equation}
\begin{eqnarray*}
%\nonumber
\varphi (A-C)=- \varphi \ln \varphi +\varphi \intnull P \ln \frac{\tp}{P}\,dE
\end{eqnarray*}
and
\begin{equation}
\begin{split}
\label{A5}
B&= \intdelta P \ln P \,dE = \intnull P^+ \ln P^+ \,dE=\varphi \intnull \tp \ln (\tp \varphi)\,dE\\
&=\varphi\left [ \intnull \tp \ln \tp \,dE +\ln \varphi \right ]
\end{split}
\end{equation}
\begin{eqnarray*}
%\nonumber
D= \intnull P^+ \ln P \,dE =\varphi \intnull \tp \ln P \,dE
\end{eqnarray*}
\begin{eqnarray*}
%\nonumber
B-D= \varphi \intnull \tp \ln \frac{\tp}{P}\,dE + \varphi \ln \varphi
\end{eqnarray*}
By adding (\ref{A4}) and (\ref{A5}) and using Gibbs' inequality for arbitrary PDF's, $\intnull dE P \ln \frac{P}{Q}\geq 0$ one finally obtains
\begin{equation}
\begin{split}
\label{A6}
\SEpunkt / (\kb w^-)=\varphi (A-C)+(B-D)
&=\varphi \left [ \intnull P \ln \frac{P}{\tp}\,dE + \intnull \tp \ln \frac{\tp}{P}\,dE\right ] \geq 0
\end{split}
\end{equation}
This proves $\SEpunkt \geq 0$, i.e. this entropy never decreases. Gibbs' inequality further implies that entropy is constant if and only if $\tp =P$, which is the condition for the stationary solution. Because of the above definition of $\tp$, this is equivalent to the condition $P(E+\Delta E)\propto P(E)$ that is satisfied only by an exponential or constant function. In order to be integrable, it must be a decreasing exponential, which proves that the stationary solution is the Maxwell-Boltzmann distribution. In summary, one can state that entropy $S_E$ always increases, $\SEpunkt > 0$, until the stationary state is reached. $\SEpunkt$ is proportional to $w^+=w^- \varphi$, which is the rate for the uptake of kinetic energy.\\
\indent The interesting one-particle phase-space entropy $S^{(1)}$ belongs not to energy, but the PDF of momentum $P^{(1)}(\textbf{p},t)=(4\pi mp)^{-1}P(E,t)$. This follows from $E=p^2/2m$ and $dp=dE\;m/p$. Integration over momenta confirms the correct normalization
\begin{eqnarray}
\int P^{(1)}(\textbf{p},t)\,d^3 p =\intnull P^{(1)}(\textbf{p},t)4\pi p^2 \,dp = \intnull P(E,t)\,dE=1
\end{eqnarray}
The connection between the two entropies is achieved by
\begin{equation}
\begin{split}
S^{(1)}&=-\kb \intnull \peins (\textbf{p},t)\ln \peins (\textbf{p},t)\,d^3 p\\
&=-\kb \intnull (4\pi mp)^{-1}P(E,t)\left [ \ln P(E,t)-\ln (4\pi m)-\ln p\right ] 4\pi p^2 \left [ dE \,m/p\right ]\\
&=-\kb \intnull P(E,t) \left [ \ln P(E,t)-\ln (4\pi m)- \frac{1}{2}\ln 2mE\right ]\, dE\\
&=S_E+\frac{1}{2} \kb \left\langle \ln E\right\rangle + const
\end{split}
\end{equation}
It seems reasonable to accept $\left\langle  \ln E\right\rangle = const$ as equivalent to energy conservation. With this assumption, one finally obtains the desired time-dependence of one-particle phase-space entropy $S^1$, namely $\dot{S}^{(1)}\geq0$. Also the entropy $S^{(1)}$ always increases, $\dot{S}^{(1)}>0$, until the stationary Maxwell-Boltzmann distribution is reached.
\pagebreak

\addtolength{\parskip}{-3mm}

% -----------------------------------------------------------
% - Literaturverzeichnis
% -----------------------------------------------------------
\section*{References}
% \begin{harvard}
\bibliographystyle{unsrtdin} %, plain, plaindin unsrt alpha alphadin
\bibliography{paper1}
% \end{harvard}

\end{document}